\documentclass[conference]{IEEEtran}
\usepackage{cite}
\usepackage{comment}
\usepackage{amsmath,amssymb,amsfonts}
\usepackage{algorithmic}
\usepackage{graphicx}
\usepackage{caption}
\usepackage{subcaption}
\usepackage{comment}
\usepackage{xfrac}
\usepackage{textcomp}
\usepackage{xcolor}
\def\BibTeX{{\rm B\kern-.05em{\sc i\kern-.025em b}\kern-.08em
    T\kern-.1667em\lower.7ex\hbox{E}\kern-.125emX}}
\begin{document}
\title{[Draft] Runtime reliability monitoring for complex fault-tolerance policies}

\author{
\IEEEauthorblockN{1\textsuperscript{st} Alessandro. Fantechi}
\IEEEauthorblockA{\textit{DINFO} \\
\textit{University of Florence}\\
Firenze, Italy \\
alessandro.fantechi@unifi.it}
\and
\IEEEauthorblockN{2\textsuperscript{nd} Gloria Gori}
\IEEEauthorblockA{\textit{DINFO} \\
\textit{University of Florence}\\
Firenze, Italy \\
gloria.gori@unifi.it}
\and
\IEEEauthorblockN{3\textsuperscript{rd} Marco Papini}
\IEEEauthorblockA{\textit{DINFO} \\
\textit{University of Florence}\\
Firenze, Italy \\
marco.papini@unifi.it}
}

\maketitle

\begin{abstract}

Reliability of complex Cyber-Physical Systems is necessary to guarantee availability and/or safety of the provided services. Diverse and complex fault tolerance policies are adopted to enhance reliability, that include a varied mix of redundancy and dynamic reconfiguration to address hardware reliability, as well as specific software reliability techniques like diversity or software rejuvenation.
These complex policies  call for flexible runtime health checks of system executions that go beyond conventional runtime monitoring of pre-programmed health conditions, also in order to minimize maintenance costs. Defining a suitable monitoring model in the application of this method in complex systems is still a challenge.
In this paper we propose a novel approach, Reliability Based Monitoring (RBM),  for a flexible runtime monitoring of reliability in complex systems, that exploits a hierarchical reliability model periodically applied to runtime diagnostics data: this allows to dynamically plan maintenance activities aimed at prevent failures. As a proof of concept, we show how to apply RBM to a $2$oo$3$ software system implementing different fault-tolerant policies.

\end{abstract}


\section{Introduction}
\label{sec:rbm_intro}

Today, computing systems often have high dependability requirements, either in terms of safety, availability, or both.
Several Cyber-Physical Systems (CPS), like distributed server farms, commercial computers and communication systems, require high availability levels, and safety is an added requirement for those employed in Safety-Critical (SC) domains, such as transportation \cite{bourbouse2016, baumgartner2001, cenelec50126-1, cenelec50126-2, arp4754a}.

In hardware components, failures are random events caused by physical phenomena, either internal or external: internal phenomena include aging effects and wear-out of components, while external phenomena include noise and interference issues \cite{cenelec50129, do254}.
Due to the stochastic nature of hardware failures, redundancy i.e., the usage of two or more identical components, is the preferred technique to increase the overall availability and safety of a CPS \cite{ieee2010_24765vocabulary}.

In software components, failures are deterministic events caused by faults in the source code, i.e., software bugs.
A software bug is triggered each time an application has a precise internal status and a given piece of source code is executed \cite{cenelec50128, do178c}.
In addition, especially in long-time running systems, error conditions such as storage fragmentation, memory leaks, unreleased memory space, round-off errors, may accumulate with time and/or system load, causing a software system to age \cite{dohi2020handbook, huang1995software}.
\cite{castelli2001proactive}.
Due to the deterministic nature of software failures, the adoption of classic redundancy techniques is not sufficient to increase safety and availability.
Formal specification and verification, as well as extensive testing coverage, is among the techniques prescribed by software safety guidelines, such as \cite{cenelec50128, do178c}, to decrease the number of residual software bugs, thus lowering the failure rate.
However, given the high costs of such techniques and the huge complexity of modern software systems, the possibility of residual bugs is non-negligible.
This is especially true when a complex system is made up of so-called Commercial off-the-shelf (COTS) components, such as Operating Systems (OS), whose reliability is not known.
Therefore, SC systems often require the adoption of redundancy with \textit{diversity}, i.e., different software is executed to avoid common-mode failures \cite{ieee2010_24765vocabulary}.

Another measure that is considered to counter software failures is \emph{software rejuvenation}, that is, restart or reset, to bring the system back to its initial state, so removing the effects of possible accumulating failures occurred in the meanwhile.
Thus, this technique restores the reliability of a software system to its initial value, and it is particularly effective to counter software aging effects. 
Software rejuvenation was introduced in the 90s \cite{dohi2020handbook} and it has been used in high availability and mission critical systems; in fact, although possible faults still remain in the software program, performing software rejuvenation occasionally or periodically reduces the probability of failures at the price of system unavailability.

These characteristics call for flexible runtime safety checks of system executions that go beyond conventional runtime monitoring of pre-programmed safety conditions, also in order to minimize maintenance costs.
The runtime selection of proper fault-tolerance policies becomes hence crucial to guarantee the required safety and availability. 

The idea put forward in this paper is that a dynamic estimation of the residual reliability can provide the optimal parameter for on-time triggering a dynamic action, improving overall reliability, while minimizing the adverse effects of such action which typically makes the system, or one of its component, temporarily unavailable.

In this paper, we present the Reliability Based Monitoring (RBM) \cite{fantechi2022wosar}, an approach that exploits both a hierarchical reliability model and diagnostics data to provide predictive diagnostics of complex systems \cite{papini_phd2021, applsci_rcps2021}.
The aim of the RBM approach is to estimate at runtime the reliability curve of the monitored system and, by leveraging this estimate, to perform prognostics, which in turn can be used to schedule maintenance interventions.
We show how RBM is conveniently applicable to schedule maintenance interventions on subsystems, with a positive and measurable impact on system reliability.

The paper is organized as follows.
In Section~\ref{sec:related} we provide a selection of related works, in Section~\ref{sec:rbm} we describe the RBM approach, in Section~\ref{sec:architecture} we introduce a simple case study, in Section~\ref{sec:example_rbm} we describe how RBM is applied to this example, in Section~\ref{sec:discussion} we discuss on the benefits and limitations of the RBM approach.
Finally, Section~\ref{sec:conclusion} presents the results and concludes the paper.

\section{Related works}
\label{sec:related}

The reliability of a system is assessed using a reliability model.
The following three main categories of models can be found in literature:
\begin{itemize}
 \item \textit{Combinatorial models}: they allow to efficiently evaluate reliability under the strong assumption of statistically independent components \cite{relAvaEng2017, handbookRAMS2018}.
 These models include Reliability Block Diagrams (RBD) \cite{moskowitz1958}, Fault Trees (FT) \cite{hixenbaugh1968},  Reliability Graphs (RG) \cite{bryant1986}.
 The expressive power varies among the different combinatorial models \cite{malhotra1994}.
 \item \textit{State-space based models}: they allow to model several dependencies among failures, including statistical, time and space dependency, at the cost of a difficult tractability due to the state-space explosion \cite{relAvaEng2017}.
 These models include Continuous-Time Markov Chains (CTMCs) \cite{stewart1994}, Stochastic Time Petri Nets (STPNs) \cite{molloy1982} and Stochastic Activity Networks (SANs) \cite{meyer1985}.
 These models have a greater expressive power than combinatorial models.
 \item \textit{Hybrid models}: they combine both combinatorial and state-space based solutions and are considered as the state-of-the-art approach to dependability evaluation \cite{relAvaEng2017}.
 Dynamic RBD (DRBD) \cite{distefano2007} and Dynamic FT (DFT) \cite{dugan1992} combine CTMC evaluation with, respectively, RBD and FT analysis.
 Hierarchical models belong to this category and exploit a layered approach: several layers, defined using different formalisms, provide a flexible reliability model and have been proposed in order to both exploit the benefits and to limit the drawbacks of combinatorial and state-space based models \cite{relAvaEng2017, papini_phd2021, rtsi2019_layered_reliability}.
\end{itemize}

Estimation of residual software reliability is a difficult task, and several methodologies have been developed \cite{wood1996, pham2006}.
These methodologies include the estimation of the number of residual bugs through the usage of software quality metrics, Non-Homogeneous Poisson Processes (NHPP) and Non-Homogeneous Continuous Time Markov Chains (NHCTMC) \cite{handbookRAMS2018, gokhale2007}.
However, these approaches produce a static reliability model, i.e., they are unable to model the time-dependent evolution of the reliability caused by the history of a monitored system.

Hardware and software reliability modeling is also treated in the literature. 
In \cite{nguyen2019_relAvalEval}, the authors show that hierarchical reliability models, which were already successfully applied to hardware systems, can also be applied to software and software/hardware systems.
The model proposed by the authors leverages RG, FT and SRN, which provide equivalent expressive power with respect to RBD and STPN \cite{malhotra1994}.

\section{Reliability Based Monitoring}
\label{sec:rbm}

Before describing the RBM approach, we have to define how the system failure rate can be modeled considering both its parts: hardware and software.
Hardware failures, which are widely studied in the literature, are typically modeled with a constant failure rate \cite{mil-hdbk-217}, even if cases of non-constant failure rate are reported, for example in electrolytic capacitors \cite{parlerCDE}.
Also software failure rate is non-constant, thus it needs to be properly modeled.

In particular, we consider the following classes of software failure modes:
\begin{itemize}
 \item \textit{Class 1}: Failures triggered by external stimuli, e.g., processing requests, that can be quite distant in time.
 The failure rate is kept constant in the time interval between two consecutive stimuli, while it is increased each time a new stimulus is received.
 This failure mode covers, e.g., lack of memory caused by issues in memory deallocation.
 \item \textit{Class 2}: Failures triggered by an excessive CPU load.
 When the CPU load increases, generally the OS has to serve a greater number of requests per unit of time.
 A higher number of OS invocations may speed up software aging.
 Furthermore, in case of real-time systems, a higher CPU load may lead to deadline miss events.
 This software failure mode is then characterized by a failure rate directly proportional to the CPU load \cite{vaidyanathan1999, wang2007}.
\end{itemize}

As already stated in Section~\ref{sec:rbm_intro}, rejuvenation is a maintenance operation aimed to restore the functionality and the reliability of a given system to a previous time instant.
When rejuvenation is executed as a preventive or predictive operation with a remote restart, the down time is usually short, while when a maintenance intervention is executed after a failure, it may require an on site intervention, and takes longer time.
To achieve the goal of maximizing the system availability while minimizing the number of rejuvenation interventions, an optimal rejuvenation policy is needed.
Our proposal is the adoption of the RBM approach, a \textit{predictive diagnostics} technique described in the following.
The RBM approach consists in the definition of a two-layered hierarchical reliability model that combines RBD and STPN and in the runtime execution of the prognostics algorithm. 
In turn, the RBM prognostics algorithm exploits an external diagnostics system and is based on: $1)$ the reconstruction of the system reliability curve; $2)$ the computation of metrics needed to assess the health of the monitored system and used to schedule predictive maintenance interventions.

The RBM approach can be applied also to those SC systems in which rejuvenation has an \textit{acceptable} impact on safety.
With \textit{acceptable} we mean that a failure against safety can occur with a probability lower than a given acceptability threshold, e.g. the Tolerable Hazard Rate (THR), which expresses the maximum number of acceptable failures against safety that can occur in one hour \cite{cenelec50126-2, arp4761}.

SC railway systems developed according to the European norms \cite{cenelec50126-2}, for example, mandate that their THR is lower than $10^{-8}$.

The subsystem under rejuvenation can be treated as \textit{failed} for the duration of the maintenance intervention.
If redundancy is used to ensure safety, the failure of a redundant subsystem within the time window of the rejuvenation may be an event against safety.
Thus, the execution of the rejuvenation has an acceptable impact on safety if the probability of such a
failure is lower than the THR.

\subsection{Model definition}
\label{subsec:rbm_model}

In the RBM  approach, the reliability model is defined once for all the analyzed systems belonging to the same type; the model will then  be instantiated for every monitored system.

We assume a system as composed of a set $I$ of independent subsystems, that in turn are structured in components: the overall set of components is denoted as $J$.
This two-level structure is intended to capture a typical  architecture made of possibly redundant hardware and the component-based software deployed over it.

It is important to analyze the overall system's architecture to determine if the system has different modes of operation.
For example, systems using redundancy techniques can operate in standard or degraded mode depending on the sets of working subsystems.
These sets determine the set $H$ of different admitted modes of operation.
The model definition is carried out through the following steps:
\begin{enumerate}
 \item \textit{Assessment} of: the $H$ modes of operation; the initial mode of operation; the information that can be collected by the diagnostics system divided in diagnostics events and data.
 \item \textit{Modeling} of the system, divided in:
 \begin{itemize}
  \item[(2a)] Decomposition into $I$ independent subsystems;
  \item[(2b)] For each  mode of operation $h \in H$, definition of the RBD model $RBD_h$;
  \item[(2c)] For each subsystem $i \in I$ , definition of the STPN model $PN_i$.
  Each $PN_i$ must contain all the critical components of the subsystem and, for all the modeled components, the transitions needed to represent the failure rates must be defined;
  \item[(2d)] For each  modeled critical component $j \in J$, definition of the failure rate function $FRF_j$ that depends on diagnostics data that will be collected by the diagnostics system.
 This function will be used at runtime to estimate the failure rates.
 \end{itemize}
 \item \textit{Initialization} of the model.
 During this step, we use the \textit{a priori} failure rates to compute: 
 \begin{itemize}
  \item[(3a)] For each  subsystem $i$, the \textit{a priori} reliability curve $R^0_i(t)$ exploiting the STPN model $PN_i$; 
  \item[(3b)] The system-level \textit{a priori} reliability curve $R^0(t)$ exploiting the RBD associated with the initial mode of operation and the previously computed reliability curves of the subsystems.
   \end{itemize}
\end{enumerate}
Note that both the RBM model and its initialization are used during the periodical execution at runtime of the RBM algorithm described in Section~\ref{subsec:rbm_implementation}.

\subsection{Runtime algorithm for prognostics}
\label{subsec:rbm_implementation}
This phase is the actual implementation of RBM prognostics and it is periodically performed with period $\delta$, for each monitored system, at time  $(n+1) \cdot \delta ~,~ \forall ~ n \geq 0$ using the diagnostics data acquired in the time interval $\left[ n \cdot \delta, ~ (n+1) \cdot \delta \right]$, where $n=0$ is the time at which the system has been switched on.
This phase is composed by the following iterative steps:
\begin{itemize}
 \item Failure rate estimation;
 \item Unconditioned subsystem reliability curve computation;
 \item Conditioned subsystem reliability curve computation;
 \item Evaluation of system reliability;
 \item Monitoring of estimated reliability.
\end{itemize}

\subsubsection{Failure rate estimation}
For each modeled component, we estimate its current failure rate using the diagnostics data acquired in the time interval $\left[ n \cdot \delta, ~ (n+1) \cdot \delta \right]$.
By feeding the diagnostics data to each function $FRF_j$, $j \in J$, we obtain the \textit{unconditioned} failure rates of the modeled components, i.e., the failure rates that would be applicable if and only if the operating conditions for all components are constant in time.
This assumption is generally false, since the operating conditions are time-dependent by their nature.

\subsubsection{Unconditioned subsystem reliability curve computation}
For each $i \in I$ subsystem, we update the STPN model $PN_i$ with the previously estimated failure rates and we evaluate the updated $PN_i$ model to retrieve the reliability curve $R'_i(t)$.

Note that, since we used the unconditioned failure rates, the reliability curves evaluated during this step are also unconditioned.

\subsubsection{Conditioned subsystem reliability curve computation}
For each $i \in I$ subsystem, we compute the conditioned reliability curve $R_i^{n+1}(t)$, i.e., the reliability curve that is built considering both the unconditioned reliability curve $R'_i(t)$ and the reliability curve obtained from the previous step $R_i^n(t)$ with $n \geq 0$.

The processing of the diagnostics events detected in time interval $\left[ n \cdot \delta, ~ (n+1) \cdot \delta \right]$ leads to the following scenarios:
\begin{itemize}
 \item \textit{Case 1:} If a subsystem has been subject to rejuvenation, its reliability is expected to increase due to the rejuvenation itself;
 \item \textit{Case 2:} If a subsystem is shut down to freeze its reliability, i.e., to keep it constant.
 Note that this possibility is applicable to hardware components only;
 \item \textit{Case 3:} In all other cases, the reliability is expected to decrease without discontinuity.
\end{itemize}

If a redundant subsystem has been subject to a rejuvenation event in time interval $\left[ n \cdot \delta, ~ (n+1) \cdot \delta \right]$ (\textit{Case 1}), the conditioned reliability $R_i^{n+1}(t)$ is computed according to~\eqref{eq:r'_rejuv}:
\begin{equation}
 R_i^{n+1}(t) = \begin{cases}
                \begin{aligned}
                 &R_i^n(t)                 & \forall ~~ & t \leq n \cdot \delta \\
                 &R'_i(t - n \cdot \delta) & \forall ~~ & t >    n \cdot \delta
                \end{aligned}
                \end{cases}
 \label{eq:r'_rejuv}
\end{equation}

If a redundant subsystem is switched off to freeze its reliability during the time interval $\left[ n \cdot \delta, ~ (n+1) \cdot \delta \right]$ (\textit{Case 2}), the conditioned reliability $R_i^{n+1}(t)$ is computed according to~\eqref{eq:r'_switchoff}:
\begin{equation}
 R_i^{n+1}(t) = \begin{cases}
                \begin{aligned}
                 &R_i^n(t)              & \forall ~~ & t \leq n \cdot \delta \\
                 &R_i^n(n \cdot \delta) & \forall ~~ & t >    n \cdot \delta
                \end{aligned}
                \end{cases}
 \label{eq:r'_switchoff}
\end{equation}

In all other cases (\textit{Case 3}), the conditioned reliability $R_i^{n+1}(t)$ is computed according to~\eqref{eq:r'_no_rejuv}:
\begin{equation}
 R_i^{n+1}(t) = \begin{cases}
                \begin{aligned}
                 &R_i^n(t)                        & \forall ~~ & t \leq n \cdot \delta \\
                 &R'_i(t - n \cdot \delta + t'_i) & \forall ~~ & t >    n \cdot \delta
                \end{aligned}
                \end{cases}
 \label{eq:r'_no_rejuv}
\end{equation}
where $t'_i$ is a constant value evaluated for each $i \in I$ subsystem and it is computed as shown in~\eqref{eq:t'_computation}:
\begin{equation}
 \begin{split}
  t'_{i} &= b \cdot \delta \mathrm{~~with~}  b \mathrm{~such~that:~}\\
  &\mathrm{~}\left| R'_i(b \cdot \delta) - R_i^n(n \cdot \delta) \right| = \min_{m} \left| R'_i(m \cdot \delta) - R_i^n(n \cdot \delta) \right|
 \end{split}
 \label{eq:t'_computation}
\end{equation}

The constant value $t'_i$ is the time instant, computed over the unconditioned curve $R'_i(t)$, that minimizes the difference with the value of the last conditioned estimate at time $n \cdot \delta$, i.e., $R_i^n(n \cdot \delta)$.

Put in other terms, we restore the reliability of a subsystem to its initial value after a rejuvenation has occurred, we keep it constant if a subsystem is currently shut down to freeze its reliability, we perform the \textit{conservation of reliability} \cite{bruneo2013}, i.e., we compute a continuous reliability curve at current time $n \cdot \delta$, otherwise.

Note that the current RBM step is executed at time $(n+1) \cdot \delta$.
When performing the conservation of reliability, we ensure the continuity of the reliability curve at time $n \cdot \delta$, i.e., we apply the conservation of reliability starting from a past time instant.
This choice is motivated as follows: the diagnostics data acquired in time interval $\left[ n \cdot \delta, ~ (n+1) \cdot \delta \right]$ and used at time $(n+1) \cdot \delta$ have an impact on the reliability starting from time $n \cdot \delta$.

\subsubsection{Evaluation of system reliability}
This step consists of the following two parts:
\begin{itemize}
 \item Selection of the RBD model $RBD_h$ to be used, based on current mode of operation.
 Starting from the RBD model selected during the previous iteration, i.e., at time $n \cdot \delta$, we process the diagnostics events acquired in time interval $\left[ n \cdot \delta, ~ (n+1) \cdot \delta \right]$ to detect if a variation of the mode of operation has occurred.
 \item Reliability evaluation for the whole system.
 This part is executed processing both the selected RBD model $RBD_h$ and the conditioned reliability curves of all the subsystems that contribute to the selected RBD.
 The output of this step is the reliability curve $R^{n+1}(t)$ conditioned to both diagnostics data and diagnostics events that have been collected since the system start-up.
\end{itemize}

We claim that this reliability curve provides a closer representation of the real reliability for past time instants since it is conditioned by diagnostics data and events, so providing a closer prediction of the real reliability curve for future time instants.

\subsubsection{Monitoring of estimated reliability}
The last step, which is the prognostics step, is performed at time $\tilde{t} = (n + 1) \cdot \delta$ using a \textit{prediction} interval of width $\Delta$.
Three different and non mutually exclusive RBM applications can be realized:
\begin{itemize}
 \item \textit{Residual reliability}.
 We evaluate the reliability at time $\tilde{t} + \Delta$ and we compare it with a fixed threshold $U$ as shown in~\eqref{eq:rbm_rel}.
 \begin{equation}
  R^{n+1}(\tilde{t} + \Delta) \leq U
  \label{eq:rbm_rel}
 \end{equation}
 If~\eqref{eq:rbm_rel} is satisfied, then the \textit{predicted} residual reliability at time $\tilde{t} + \Delta$ is considered too low, hence a corrective action has to be taken.
 We propose the adoption of $U = e^{-1}$, i.e., the reliability value of a system with constant failure rate evaluated at time equal to its Mean Time To Failure (MTTF).
 
 \item \textit{Probability of failure}.
 We compute the probability that the system will fail before time $\tilde{t} + \Delta$ given that it is correctly operating at time $\tilde{t}$ as shown in~\eqref{eq:rbm_pfail_formula}:
 \begin{equation}
  P_{SL}(t \in [\tilde{t}, \tilde{t} + \Delta]) = \frac{R^{n+1}(\tilde{t}) - R^{n+1}(\tilde{t} + \Delta)}{R^{n+1}(\tilde{t})}
  \label{eq:rbm_pfail_formula}
 \end{equation}
 where $P_{SL}(t)$ is the probability of system loss at time $t$.
 We compare it with a system dependent threshold $V$ and, if the probability of failure is greater than the threshold, a corrective action has to be taken.
 
 \item \textit{Remaining Useful Life (RUL)}.
 We compute the time of End of Life (EoL) $t_{eol}$ according to Eq.~\ref{eq:t_eol}:
 \begin{equation}
  R^{n+1}(t_{eol}) = e^{-1}
  \label{eq:t_eol}
 \end{equation}
 By definition, the RUL is the difference between the EoL and the current time: $RUL = t_{eol} - \tilde{t}$.
\end{itemize}

The choice of parameter $\Delta$ and of thresholds $U$ and $V$ is system dependent.
The RBM approach can be extended by defining multiple values for the prediction interval and the thresholds.
If the number of monitored prediction intervals is equal to $p$, we will have $1 \leq q \leq p$ tuples $\langle \Delta_q, U_q, V_q \rangle$ over which RBM will be applied.

\section{Proof of concept - System description}
\label{sec:architecture}

\begin{figure}[ht]
 \centering
 \includegraphics[width=0.7\columnwidth]{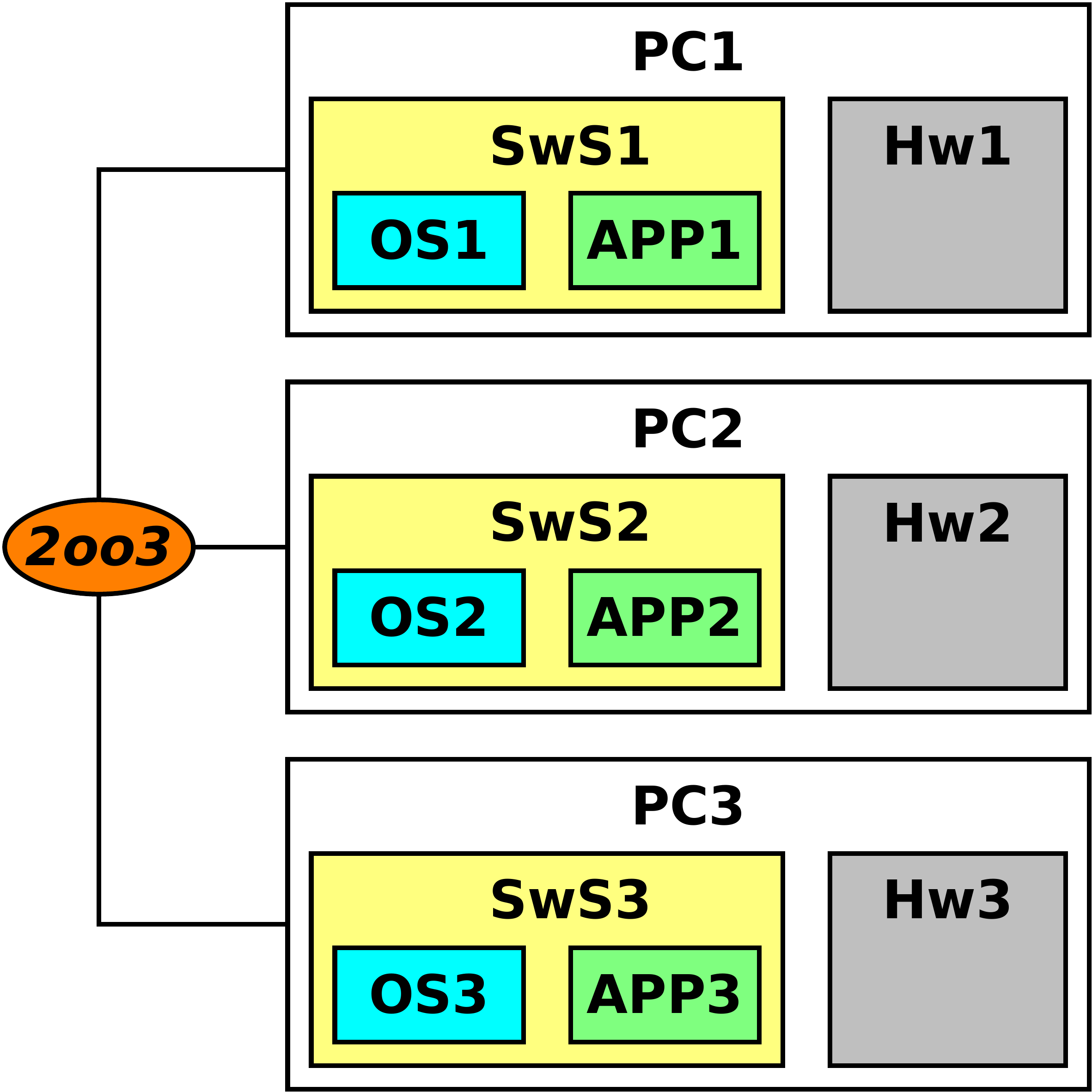}
 \caption{Architecture of the modeled CPS}
 \label{fig:architecture}
\end{figure}

In this section, we describe a system used to realize a proof of concept for the application of the RBM approach.
We consider a CPS that uses $2$oo$3$ hardware redundancy, i.e., it is composed by three identical industrial computers.
Each industrial PC executes a Software System (SwS) contributing to the computation in a $2$oo$3$ software architecture with software diversity.
Each SwS hosts an OS and it executes an Application that awaits external stimuli to execute the proper control algorithm.
The architecture is shown in Figure~\ref{fig:architecture}.

The presented $2$oo$3$ system provides a fault-tolerant architecture that covers both hardware and software failures.
To simplify the implementation of this proof of concept, we consider only software-related failures.

Each SwS has three different states of operation, namely \textit{init}, \textit{working} and \textit{failed}.
The \textit{init} state includes both the OS and Application start-up: while in this state, the SwS cannot cooperate to the overall system.
The operations required by this state are carried out in a stochastic time distributed according to the Mean Time To Init (MTTI).
While in \textit{working} state, the SwS is correctly executing both the OS and the Application and it is cooperating to the overall CPS.
The time interval in which the SwS is in \textit{working} state is distributed according to a stochastic time.
Finally, in \textit{failed} state, the SwS has suffered a failure either due to the OS or to the Application and it cannot cooperate to the overall system.
The stochastic time needed to repair the failed SwS is distributed according to the Mean Time To Repair (MTTR).

Since an SwS can contribute to the $2$oo$3$ system if and only if it is in \textit{working} state, we can reduce the number of modeled states to the following two: \textit{available} and \textit{unavailable}.

Software failures can occur in a single SwS either due to the OS or to the Application and cause the affected SwS to become unavailable.
In this case, one of the following scenarios can occur:
\begin{itemize}
 \item \textit{Scenario A}: if the other two SwSs are correctly operating, then the CPS is available.
 \item \textit{Scenario B}: otherwise, the number of working SwSs is lower than $2$ and the CPS is unavailable.
\end{itemize}

In both cases, a maintenance intervention is planned to restore the failed SwS: this maintenance operation will be carried out in a stochastic time according to the sum of both MTTR and MTTI.
To minimize the probability of occurrence of \textit{Scenario B}, predictive software rejuvenation can be scheduled and performed, leveraging the RBM approach. 
Note that, in case of predictive rejuvenation operations, the down-time of the impacted SwS is generally lower than the one in case of failure. 
In fact, while in the latter case the stochastic down-time is distributed according to the sum of both MTTR and MTTI, in the former only MTTI is involved.
We discussed more in depth the benefits provided by the RBM approach to systems implementing software rejuvenation in \cite{fantechi2022wosar}.

\section{Proof of concept - RBM approach}
\label{sec:example_rbm}

In this section, we show how the RBM approach can be applied to model and monitor the system described in Section~\ref{sec:architecture}.
In particular, in Section~\ref{subsec:example_rbm_model} we define the RBM model, while in Section~\ref{subsec:example_rbm_prognostics} we show how the prognostics provided by the periodic execution of the RBM algorithm can be exploited to increase the availability of the monitored system.

\subsection{RBM model definition}
\label{subsec:example_rbm_model}

In the following, 
we use the numbers in parenthesis to refer to the corresponding steps enumerated in Section~\ref{subsec:rbm_model}.

$(1)$ \textit{Assessment}.
The following two modes of operation exist: \textit{operating}, in which all SwSs are available; \textit{degraded}, in which one SwS is failed.
The initial mode of operation is \textit{operating}: indeed, the start-up time of the three systems may slightly differ, but it is expected to be similar.
Diagnostics events include the power-on and the shut-down of an SwS, either due to rejuvenation or to a failure, together with the identifier of the affected SwS.
If one of the three SwSs encounters an issue and it is unable to complete the start-up, the diagnostic system detects this issue after a timeout and it raises the proper diagnostics event, thus causing the model to evolve to \textit{degraded} mode of operation.
Finally, the diagnostics system acquires, for each SwS, both the number of external stimuli since its power-on and the average CPU load.

$(2)$ \textit{Modeling} of the system, subdivided into the following steps.

$(2\mathrm{a})$ The system is divided into three independent subsystems, i.e., the three industrial PCs with their SwSs: as already stated, we focus only on the SwSs.

\begin{figure}[ht]
 \centering
 \begin{subfigure}[t]{\columnwidth}
  \centering
  \includegraphics[width=0.6\columnwidth]{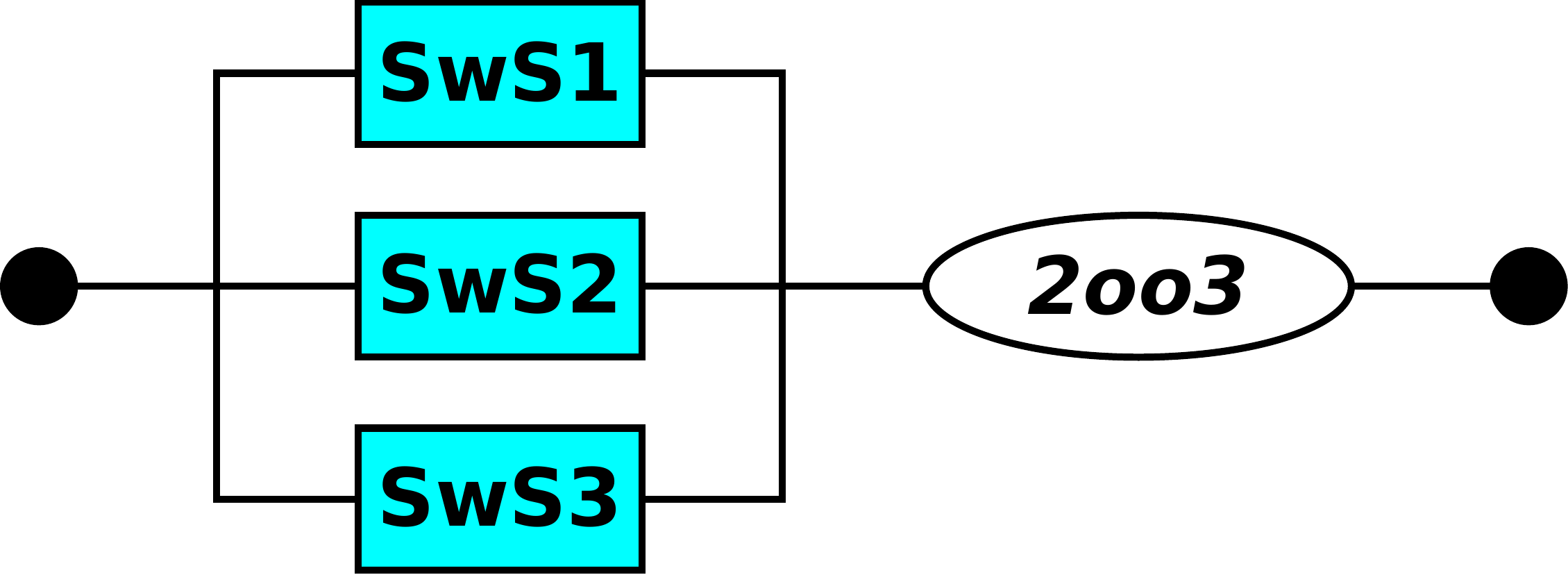}
  \caption{RBD of \textit{operating} mode}
  \label{subfig:rbd_operating}
 \end{subfigure}
 \hfill
 \begin{subfigure}[t]{\columnwidth}
  \centering
  \includegraphics[width=0.6\columnwidth]{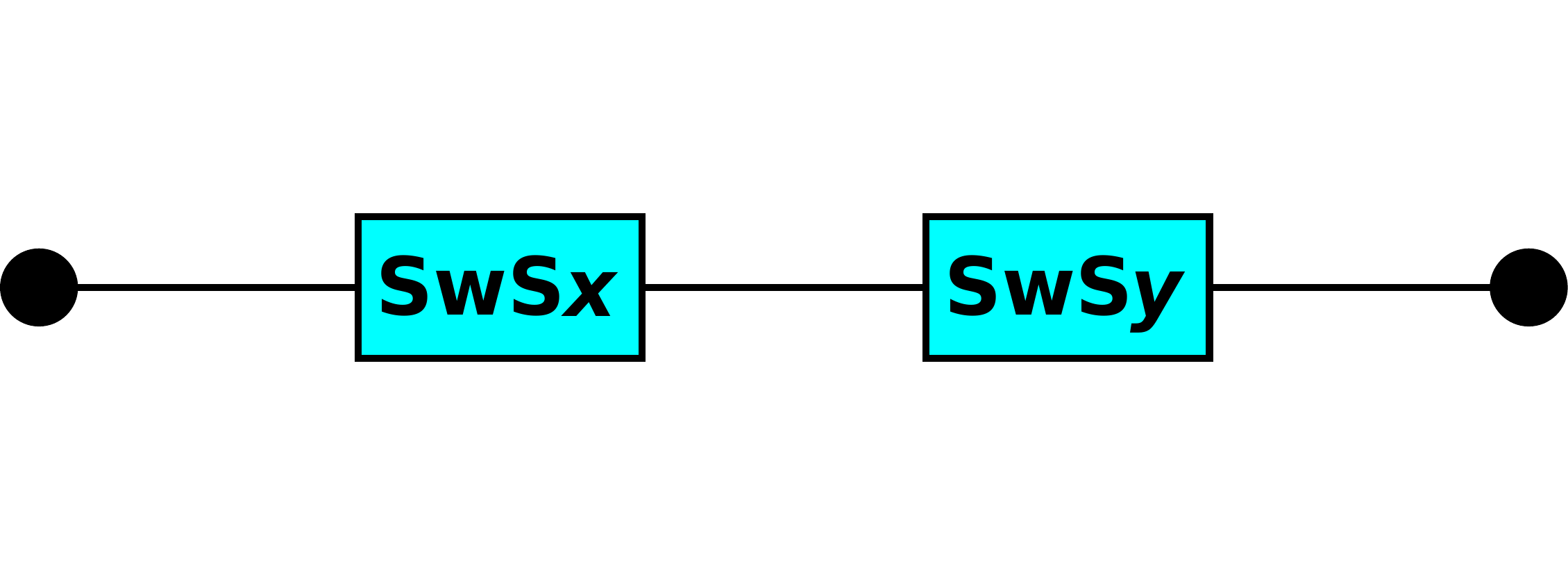}
  \caption{RBD of \textit{degraded} mode}
  \label{subfig:rbd_degraded}
 \end{subfigure}
 \caption{RBDs of analyzed system}
 \label{fig:rbd}
\end{figure}

$(2\mathrm{b})$ Two RBD models are defined, the $2$oo$3$ block shown in Figure~\ref{subfig:rbd_operating} for \textit{operating} mode, and the series block shown in Figure~\ref{subfig:rbd_degraded} for \textit{degraded} mode, with $x, y \in [1 ... 3], x \neq y$.

\begin{figure}[ht]
 \centering
 \includegraphics[width=0.6\columnwidth]{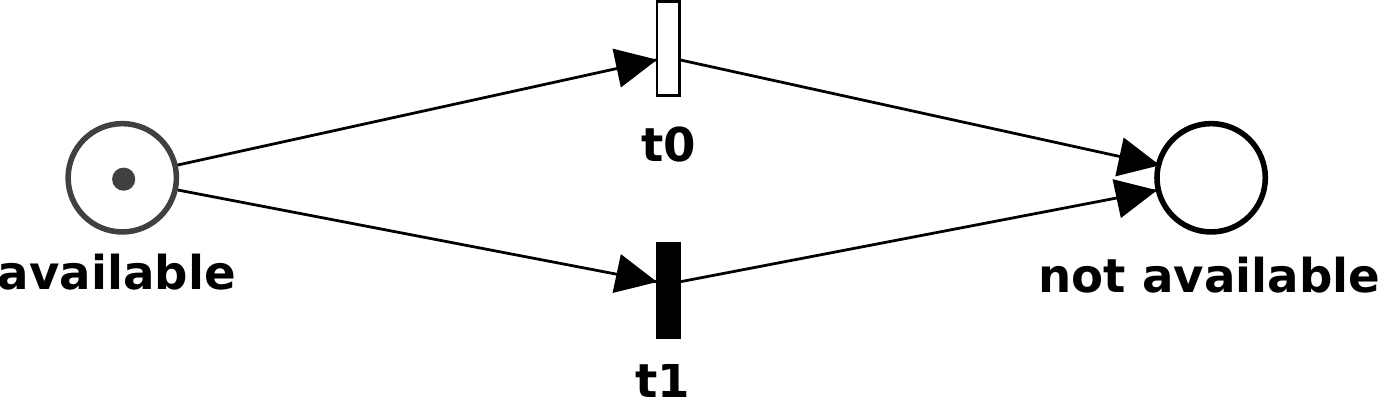}
 \caption{STPN model for failures of an SwS}
 \label{fig:pn}
\end{figure}

$(2\mathrm{c})$ For what concerns the STPN, we model failures due to both  the OS and  the Application according to the classes defined in Section~\ref{sec:rbm}.
We assume that failures due to the Application belong to Class 1 and they are mainly due to lack of memory.
This implies that, between two subsequent stimuli, the failure rate is constant, while it increases linearly when a new stimulus is received.
We assume that the OS fails when a Poisson process reaches the $k$-th fault, with faults distributed as exponential variables with rate $\lambda$.
Faults of the OS belong to Class 2 and include, among others, issues in interrupt management and in system calls management.
Hence, failures of the OS can be modeled using the Erlang distribution.
The STPN used to model the failures of a given SwS is shown in Figure~\ref{fig:pn}, where \textbf{t0} is an exponential transition modeling the Application failures and \textbf{t1} is the Erlang transition modeling the OS failures.

$(2\mathrm{d})$ 
For what concerns the failure rate functions, we model the Application failures using $r_{app}(t) = \alpha + s(t) \cdot \beta$, where: $r_{app}(t)$ and $s(t)$ are respectively the failure rate and the number of external stimuli at time $t$; $\alpha$ and $\beta$ are two constants that depend on the Application.
The OS failures are distributed according to the Erlang distribution $Erl(k(t), \lambda(t))$ with $\lambda(t) \propto L_{cpu}(t)$, where $L_{cpu}(t)$ is the average CPU load at time $t$.
The monotonic non increasing function $k(t)$ models the number of faults, at time $t$, that can occur before the OS failure.
This function is characterized by its starting value $k(0) = k_0$ and its value at time $(n+1) \cdot \delta$ is updated using the diagnostics data acquired in time interval $\left[ n \cdot \delta, ~ (n+1) \cdot \delta \right]$.
Note that, due to the adoption of software diversity, the model parameters $\alpha$, $\beta$, $k_0$ and function $\lambda(t)$ are different for each SwS.

\begin{figure*}[ht]
 \centering
 \includegraphics[width=0.9\textwidth]{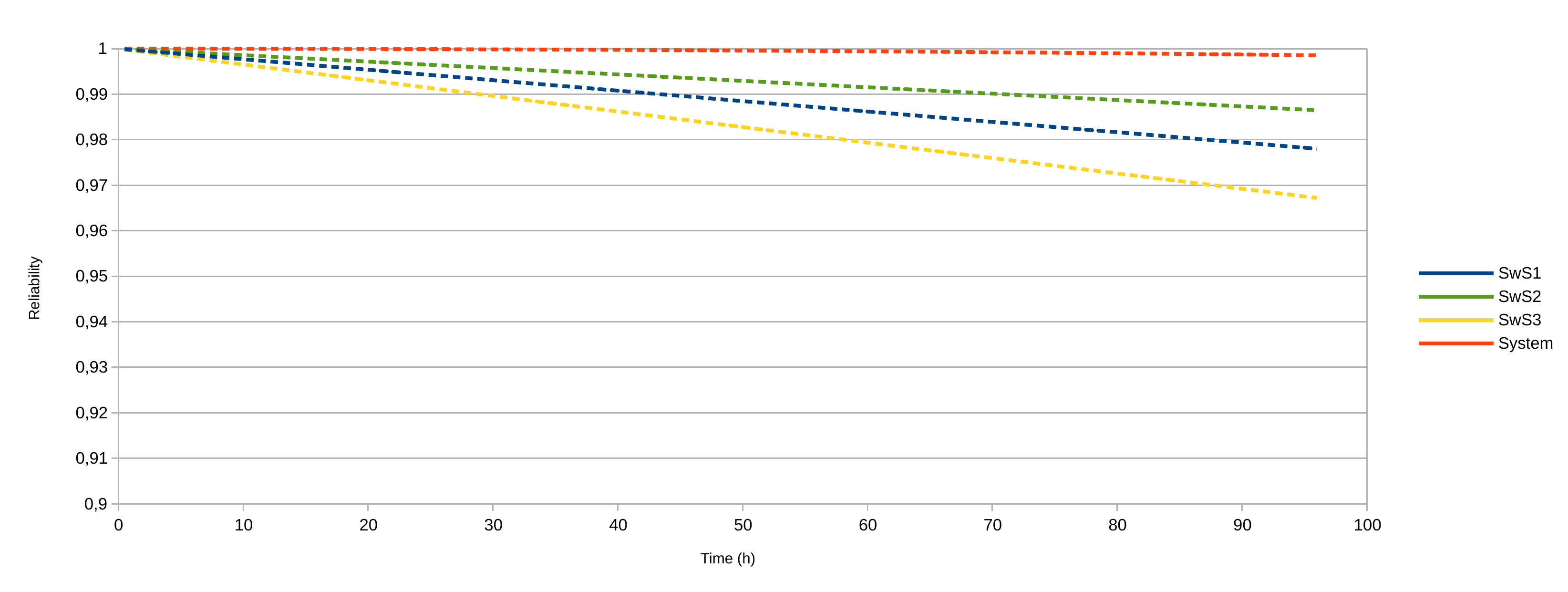}
 \caption{Initialization of RBM model}
 \label{fig:r0}
\end{figure*}

$(3)$ \textit{Initialization} of the RBM model.
By feeding the \textit{a priori} failure rate functions of all components to the three STPN models we compute the \textit{a priori} reliability curves of all subsystems and, by applying the RBD associated with the initial mode of operation, we compute the system-level \textit{a priori} reliability curve.
The \textit{a priori} reliability curves computed during the execution of this step are shown in Figure~\ref{fig:r0}.
In particular, the blue, green and yellow curves show respectively the reliability of SwS$1$, SwS$2$ and SwS$3$, while the red one shows the reliability of the entire software system.
Note that the initialization step provides only unconditioned reliability curves, since no diagnostics data and event has been collected yet.
The computation of the reliability curves has been performed by means of a tool set composed by \emph{ORIS} tool for the modeling and evaluation of STPNs \cite{paolieri2019} and \emph{librbd} for the evaluation of RBDs \cite{applsci_rcps2021}.

To successfully schedule predictive rejuvenation on the available SwSs, we apply the RBM approach both to the entire system and to the single SwSs.
Predictive rejuvenation can be scheduled only when the system is operating in $2$oo$3$ mode according to the following conditions:
\begin{itemize}
 \item When the RBM approach applied to a single SwS triggers a diagnostics alarm, we schedule a rejuvenation for the selected SwS;
 \item When the RBM approach applied to the entire system triggers a diagnostics alarm, we schedule a rejuvenation for the SwS with lowest reliability.
\end{itemize}

The stochastic parameters that have been defined during the modeling phase, including both the MTTI and the MTTR, are summarized in Table~\ref{tab:parameters}.
These parameters should depend on the actual system implementation.
In our proof of concept, these parameters, although reasonable, do not derive from such analysis.
\begin{table}[htbp]
\caption{Stochastic parameters of the RBM model}
\begin{center}
\begin{tabular}{|l|r|r|r|}
\cline{2-4}
\multicolumn{1}{l|}{} & \multicolumn{1}{l|}{SwS1} & \multicolumn{1}{l|}{SwS2} & \multicolumn{1}{l|}{SwS3} \\ \hline
$\alpha$ $(\mathrm{h}^{-1})$ & ${1}/{8640}$ & ${1}/{14112}$ & ${1}/{5760}$ \\ \hline
$\beta$ $(\mathrm{h}^{-1})$ & ${1}/{8013}$ & ${1}/{15158}$ & ${1}/{21936}$ \\ \hline
$\lambda^1$ $(\mathrm{h}^{-1})$ & ${1}/{1267}$ & ${1}/{4637}$ & ${1}/{3326}$ \\ \hline
$\lambda^2$ $(\mathrm{h}^{-1})$ & ${1}/{1901}$ & ${1}/{1987}$ & ${1}/{2722}$ \\ \hline
$k_0$ & {7} & {5} & {11} \\ \hline
$\lambda(t)$ & \multicolumn{ 3}{c|}{$\lambda^1 + \lambda^2 \cdot L_{cpu}(t)$} \\ \hline
$\delta$ & \multicolumn{ 3}{c|}{30 minutes} \\ \hline
$\Delta_q$ & \multicolumn{ 3}{c|}{$q$ days, $1 \leq q \leq 3$} \\ \hline
$U_q$ & \multicolumn{ 3}{c|}{$U_q=U=e^{-1}$, $1 \leq q \leq 3$} \\ \hline
MTTI & \multicolumn{ 3}{c|}{3 minutes} \\ \hline
MTTR & \multicolumn{ 3}{c|}{40 minutes} \\ \hline
\end{tabular}
\end{center}
\label{tab:parameters}
\end{table}

The interval width $\delta$ has been set to $30$ minutes.
This choice allows  both to collect enough diagnostics data and events and to keep a simple analysis of the described system.

Finally, the prediction interval $\Delta$ has to be higher than the sum of the MTTI and the MTTR since we want to avoid that an additional SwS failure occurs when a rejuvenation is in progress.
We apply the RBM approach using three different tuples $\langle \Delta_q, U_q \rangle, 1 \leq q \leq 3$ by posing $\Delta_q = q$ days and $U_q = U = e^{-1}$, i.e., the residual reliability evaluated at $t$ equal to MTTF assuming constant failure rate.
This choice allows us to schedule rejuvenation events with the following criteria:
\begin{itemize}
 \item When the evaluation at $\tilde{t} + \Delta_3$ is lower than $U$, a low-priority alarm is triggered and rejuvenation is planned during nightly hours since the system load is expected to be minimum during this period of time;
 \item When the evaluation at $\tilde{t} + \Delta_2$ is lower than $U$, a medium-priority alarm is triggered and rejuvenation is planned within twelve hours, trying to choose a period with low to medium load;
 \item When the evaluation at $\tilde{t} + \Delta_1$ is lower than $U$, an high-priority alarm is triggered and rejuvenation has to be performed as soon as possible.
\end{itemize}
Some CPSs are known to have predetermined time periods of low system load, while for other systems this \textit{a priori} knowledge may not be available.
Nonetheless, it is possible to estimate time periods of low system loads through an analysis of both current and historical diagnostics data.

The RBM approach can cover both  hardware and software subsystems.
We can model the system-level reliability by using a $2$oo$3$ block composed of the series of one industrial PC with its respective SwS.
This model is feasible since, in general, hardware-related failures are statistically independent from software-related ones \cite{iannino1990, lyu1996}.

\subsection{Prognostics with RBM algorithm}
\label{subsec:example_rbm_prognostics}

To apply the RBM prognostics, we define a synthetic diagnostics dataset that allows us to reach critical phases in short times while still being compatible with a real usage profile.
To this purpose, the built dataset simulates heavy stress conditions for the monitored SwSs, both in terms of external stimuli received and in terms of CPU load.

\begin{figure*}[ht]
 \centering
 \includegraphics[width=0.9\textwidth]{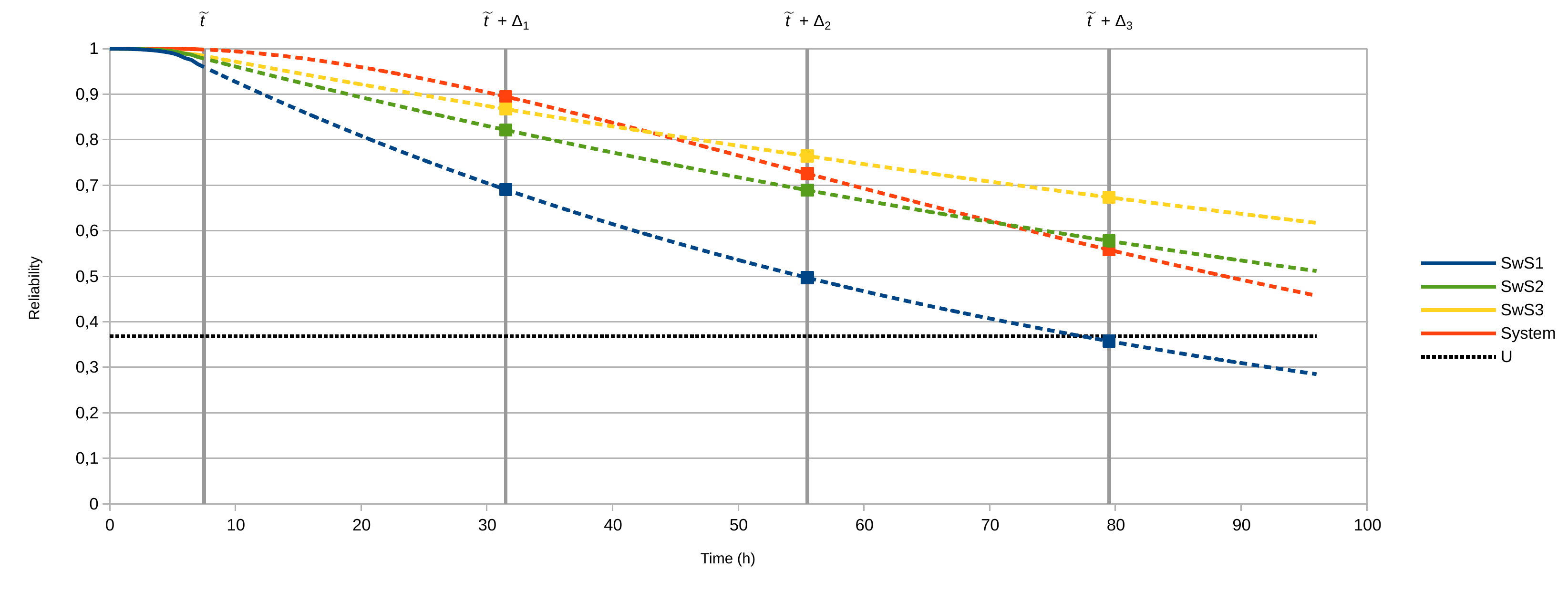}
 \caption{RBM evaluation after $7$:$30$ hours}
 \label{fig:r14}
\end{figure*}

\begin{figure*}[ht]
 \centering
 \includegraphics[width=0.9\textwidth]{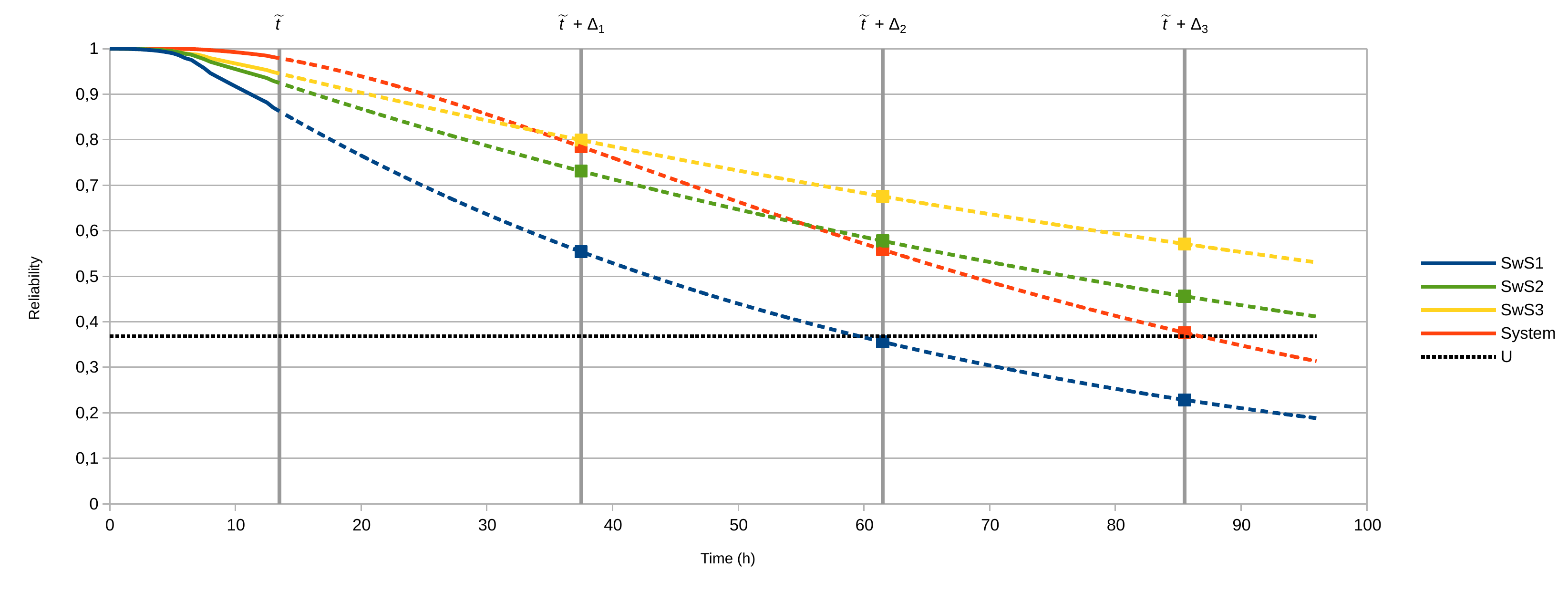}
 \caption{RBM evaluation after $13$:$30$ hours}
 \label{fig:r26}
\end{figure*}

\begin{figure*}[ht]
 \centering
 \includegraphics[width=0.9\textwidth]{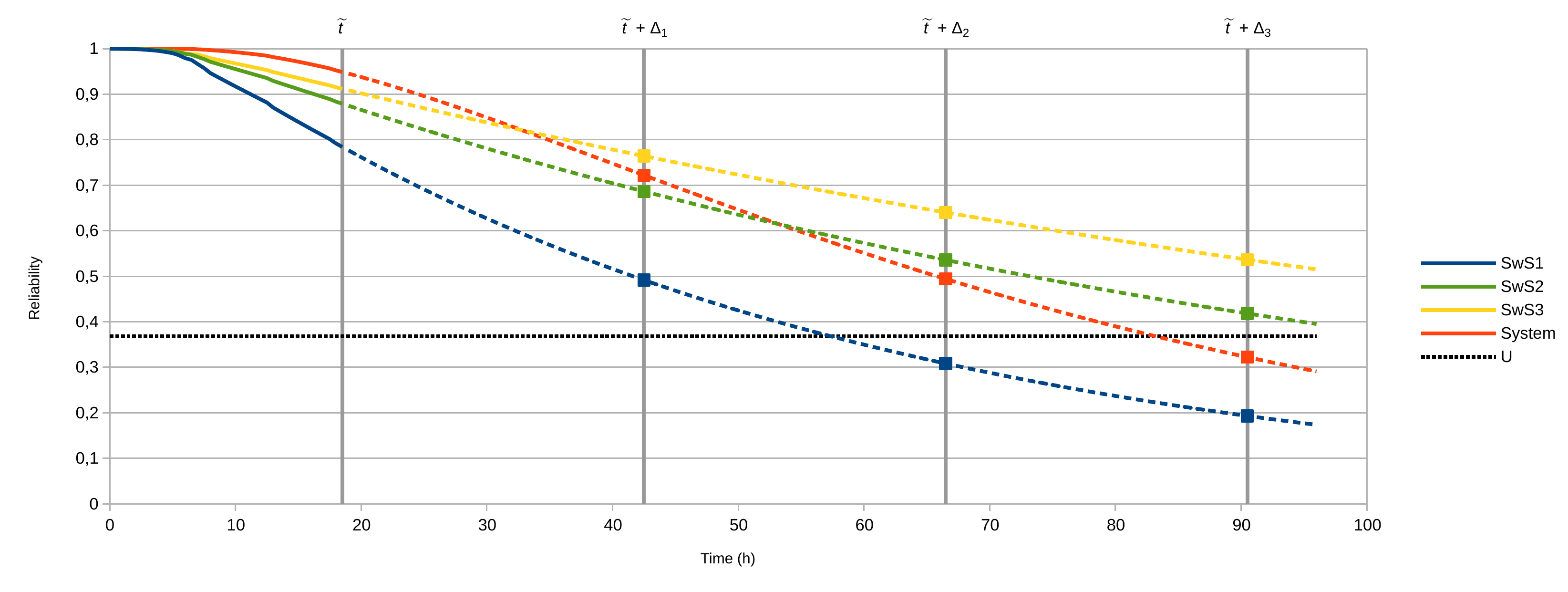}
 \caption{RBM evaluation after $18$:$30$ hours}
 \label{fig:r36}
\end{figure*}

\begin{figure*}[ht]
 \centering
 \includegraphics[width=0.9\textwidth]{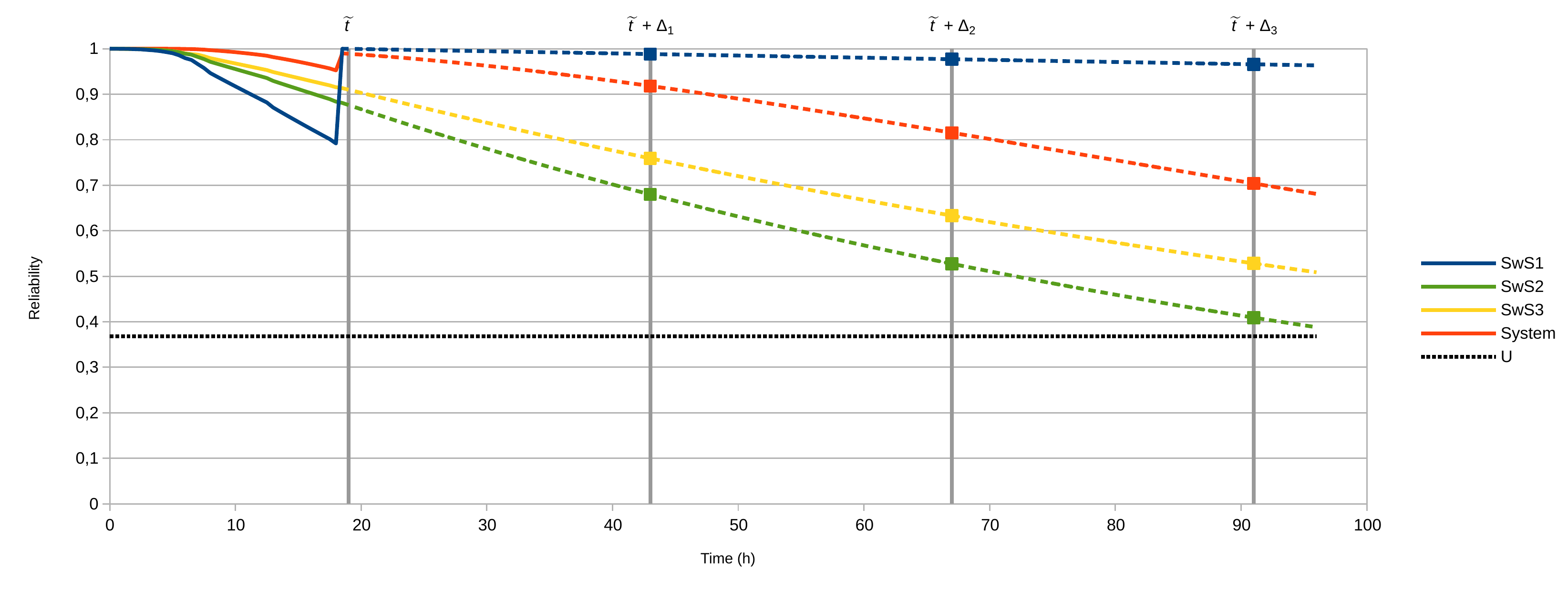}
 \caption{RBM evaluation after $19$:$00$ hours.}
 \label{fig:r37}
\end{figure*}

Figures~\ref{fig:r14}-\ref{fig:r37} show the results of the RBM approach applied to different time instants.
Dashed lines show the \textit{predicted} reliability curve, i.e., the one estimated by using the available diagnostics data and events, while continuous lines show the \textit{observed} reliability curve, i.e., the one computed until the current time.
The black dashed line represents the threshold $U = e^{-1}$ used to raise the diagnostics alarms.
The four gray vertical bars identify the key instants of each result, i.e., the current instant $\tilde{t}$ and the three prediction instants $\tilde{t} + \Delta_{q}$ at which the reliability curves are evaluated for the management of diagnostics alarms.
Finally, the squares drawn on the prediction curves are used to remark the reliability evaluated at the three prediction instants $\tilde{t} + \Delta_{q}$.

Figure~\ref{fig:r14} shows the reliability curves after $7$:$30$ hours.
If we compare it with the \textit{a priori} reliability curves shown in Figure~\ref{fig:r0}, we can see that the three SwSs are affected by degradation due to the heavy stress conditions.
Furthermore, we observe that the reliability curves of all SwSs are very different due to the adoption of software diversity.
Focusing on SwS$1$, which is the most degraded subsystem, we observe that $R_{1}(\tilde{t} + \Delta_{3}) < U$.
Put in other terms, the predicted reliability with a forecast interval of three days is lower than the alarm threshold, hence a low-priority diagnostics alarm is raised.
As already stated in Section~\ref{subsec:example_rbm_model}, this alarm triggers a rejuvenation request for SwS$1$.
The rejuvenation is planned within $18$:$00$ hours, i.e., during nightly hours with expected low load on the overall system.
It is important to remark that, even if a diagnostics alarm has been triggered for SwS$1$, the RBM approach continues to monitor its reliability for prognostics: this allows to bring forward the already requested rejuvenation if further degradation is detected.

Figure~\ref{fig:r26} shows the reliability after $13$:$30$ hours.
We observe that SwS$1$ is suffering additional degradation due to the severe operating conditions: this deduction becomes obvious when we note that $R_{1}(\tilde{t} + \Delta_{2}) < U$.
The predicted reliability with a forecast interval of two days has become lower than the alarm threshold, hence an additional medium-priority diagnostics alarm is raised.
This alarm triggers a new rejuvenation request for SwS$1$, which forces to reschedule the maintenance intervention within $6$:$00$ hours.

Figure~\ref{fig:r36} shows the reliability curve after $18$:$30$ hours.
We observe that the degradation of SwS$1$ has slowed down but it has not stopped.
Furthermore, we can observe that also the reliability of SwS$2$ is suffering from degradation, albeit $R_{2}(\tilde{t} + \Delta_{3}) > U$.
The probability of observing a failure of SwS$1$ within three days is almost $80\%$, while the probability of having a failure of SwS$2$ within the same time is approximately $58\%$.
The predicted reliability curve for the overall system is then rapidly decreasing.
This causes that $R(\tilde{t} + \Delta_{3}) < U$, i.e., the probability of system loss within three days is too high.
An additional system-level and low-priority diagnostics alarm is raised.
As described in Section~\ref{subsec:example_rbm_model}, this alarm forces the evaluation of the remaining reliability of all subsystems to select the SwS with lower reliability.
SwS$1$ is selected and a new low-priority diagnostics alarm is triggered: since the same alarm is already active, this condition is discarded without further action.

Finally, Figure~\ref{fig:r37} shows the reliability curve after $19$:$00$ hours, i.e., after the predictive rejuvenation on SwS$1$ has been carried out.
First of all, we note that the rejuvenation event on SwS$1$ at time $\tilde{t}$ has restored its reliability to the initial value $1$.
The direct consequences of the rejuvenation on SwS$1$ can be observed also at system-level: in fact, the overall system reliability at time $\tilde{t}$ has increased by almost $4\%$.
The beneficial effects of rejuvenation are even higher when we observe the predicted reliability at time $\tilde{t} + \Delta_{3}$, for which the reliability since the system start-up has doubled from slightly less than $33\%$ to almost $70\%$.
From the analysis of these results, we deduce that the application of prognostics exploiting the RBM approach allows to increment both  system-level reliability and availability.
\section{Discussion on RBM approach}
\label{sec:discussion}

In this section, we discuss the advantages and limitation of the RBM approach emerged from its application to the example system.

In particular, we can highlight the following  advantages:
\begin{itemize}
 \item The usage of a hierarchical reliability model allows to represent complex architectures, i.e., composed by several components and implementing static measures, e.g., redundancy and software diversity.
 The adoption of hierarchical reliability models is extremely useful when a subsystem is updated or retrofitted, e.g., OS or Application update for bug fixing: in such case, only the model of the updated subsystem, including its failure rate functions, has to be updated.
 Furthermore, the possibility to use several distribution functions for the definition of the subsystems models provides a high flexibility to the overall reliability model;
 \item Traditional approaches could be used to guide the definition of failure rate functions $FRF_j$, in particular to estimate their parameters.
 Furthermore, the runtime data needed to implement the RBM prognostics can be collected using automated log analysis techniques \cite{he2021};
 \item The RBM approach can be exploited to dynamically schedule predictive maintenance interventions with a cost-effective policy.
 The aim of these predictive maintenance interventions is to keep a high level of system availability.
 To minimize the negative effects of a system downtime, which can still occur with a residual probability, maintenance interventions are scheduled during a period of low system load.
 \item The RBM approach supports the following three different metrics: $1)$ estimation of the Residual Reliability; $2)$ computation of the Probability of Failure; $3)$ estimation of the Remaining Useful Life.
 These metrics can be combined to refine the management of diagnostics alarms needed to schedule the rejuvenation;
 \item The RBM approach can cover both hardware and software subsystems.
 We can model the system-level reliability by using a top-layer combinatorial model which combines both hardware and software subsystems.
 This model is feasible since, in general, hardware-related failures are statistically independent from software-related ones \cite{iannino1990, lyu1996};
 \item The model is dynamic, i.e., it evolves on the basis of diagnostics events and data periodically acquired at runtime.
The current mode of operation, which can be extrapolated from diagnostics events, has a direct impact on the reliability model in use, while the FRFs used to model the transitions of the stochastic models are updated at runtime using the currently acquired diagnostics data;
 \item Several systems that share the same architecture can be monitored, at the same time, using the same RBM model.
 In particular, these systems naturally share: $RBD_h$ and $PN_i$ models; failure rate functions $FRF_j$ used to adjust the failure rates.
 The cost of the model definition are amortized between all monitored systems.
 The cost of acquiring diagnostics events and data, to dynamically update the RBM models and to perform prognostics is, on the other hand, linear with the number of monitored systems.
\end{itemize}

On the other hand, the RBM approach suffers from the following issues:
\begin{itemize}
 \item The decomposition of a system into independent subsystems is, in general, a non-trivial task.
 Furthermore, in some conditions, this decomposition may not be possible;
 \item The complexity of the RBM model may lead to great effort needed to define it.
 In particular, the effort is divided between the definition of the hierarchical reliability model and the definition of the failure rate functions used to adjust the failure rates;
 \item The application of the approach is currently not automated.
 A tool that supports all the different phases, from the hierarchical model definition to the automatic execution of prognostics, has to be developed.
\end{itemize}

\section{Conclusions}
\label{sec:conclusion}

In this paper, we presented Reliability Based Monitoring (RBM), a novel approach to predictive maintenance exploiting diagnostics events and data to design a cost-effective maintenance policy while monitoring the system health condition.
We have discussed, by means of an example, how RBM can be used to improve system reliability implementing different fault-tolerance policies.
The reliability estimation of the monitored system is periodically updated exploiting diagnostics events and data and, by using this estimation, diagnostics alarms are managed.
The maintenance interventions are then scheduled according to diagnostics alarms, trying to prioritize a time interval with expected low system load.
This policy minimizes both the probability of having a system loss during maintenance, since the probability of failure is expected to lower when the system load decreases, and the negative impacts of a possible system failure, since the impact of system unavailability is also at its minimum.
As reported in \cite{fantechi2022wosar}, the monitoring nature of the approach, together with the benefits provided by the software rejuvenation, is an effective combination against software aging effects.

Several developments are needed: the benefits of the RBM approach are promising, but a full validation exploiting an industrial case study is necessary.
An additional step of future work is the development of an integrated tool to support all the phases of the approach, starting from the definition of the hierarchical model to the automated execution of the prognostics algorithms.

\bibliographystyle{plain}
\bibliography{main}

\begin{thebibliography}{10}

\bibitem{baumgartner2001}
J.~P. Baumgartner.
\newblock {\em {Prices and Costs in the Railway Sector}}.
\newblock \'{E}cole Polytechnique F\'{e}d\'{e}rale de Lausanne, 2001.

\bibitem{bourbouse2016}
S.~{Bourbouse}, J.~{Blanquart}, J.~F. {Gajewski}, and C.~{Lahorgue}.
\newblock {Evaluation of EEE Reliability Prediction Models for Space
  Applications}.
\newblock In {\em Proc. of DSN-W 2016}, pages 218--221, 2016.

\bibitem{bruneo2013}
D.~Bruneo, S.~Distefano, F.~Longo, A.~Puliafito, and M.~Scarpa.
\newblock Workload-based software rejuvenation in cloud systems.
\newblock {\em IEEE Transactions on Computers}, 62(6):1072--1085, 2013.

\bibitem{bryant1986}
R.~E. Bryant.
\newblock {Graph-Based Algorithms for Boolean Function Manipulation}.
\newblock {\em IEEE Transactions on Computers}, C-35(8):677--691, 1986.

\bibitem{applsci_rcps2021}
L.~Carnevali, L.~Ciani, A.~Fantechi, G.~Gori, and M.~Papini.
\newblock {An Efficient Library for Reliability Block Diagram Evaluation}.
\newblock {\em Applied Sciences}, 11(9), 2021.

\bibitem{rtsi2019_layered_reliability}
L.~{Carnevali}, L.~{Ciani}, A.~{Fantechi}, and M.~{Papini}.
\newblock A novel layered approach to evaluate reliability of complex systems.
\newblock In {\em Proc. of RTSI 2019}, pages 291--295, 2019.

\bibitem{castelli2001proactive}
V.~Castelli, R.~E. Harper, P.~Heidelberger, S.~W. Hunter, K.~S. Trivedi,
  Kalyanaraman Vaidyanathan, and William~P Zeggert.
\newblock Proactive management of software aging.
\newblock {\em IBM Journal of Research and Development}, 45(2):311--332, 2001.

\bibitem{cenelec50129}
{CENELEC}.
\newblock {EN-50129: Railway Applications -- Safety related electronic railway
  control and protection systems}, 1994.

\bibitem{cenelec50128}
{CENELEC}.
\newblock {EN-50128:2011: Railway Applications -- Communication, signalling and
  processing systems -- Software for railway control and protection systems},
  2011.

\bibitem{cenelec50126-1}
{CENELEC}.
\newblock {EN 50126-1: Railway Applications -- The Specification and
  Demonstration of Reliability, Availability, Maintainability and Safety (RAMS)
  - Part 1: Generic RAMS Process}, 2017.

\bibitem{cenelec50126-2}
{CENELEC}.
\newblock {EN 50126-2: Railway Applications -- The Specification and
  Demonstration of Reliability, Availability, Maintainability and Safety (RAMS)
  -- Part 2: Systems Approach to Safety}, 2017.

\bibitem{distefano2007}
S.~Distefano and A.~Puliafito.
\newblock Dynamic reliability block diagrams: Overview of a methodology.
\newblock {\em Proc. of ESREL 2007}, 2, 01 2007.

\bibitem{dohi2020handbook}
T.~Dohi, K.~S. Trivedi, and A.~Avritzer.
\newblock {\em Handbook of Software Aging and Rejuvenation: Fundamentals,
  Methods, Applications, and Future Directions}.
\newblock World Scientific, 2020.

\bibitem{dugan1992}
J.~B. {Dugan}, S.~J. {Bavuso}, and M.~A. {Boyd}.
\newblock Dynamic fault-tree models for fault-tolerant computer systems.
\newblock {\em IEEE Transactions on Reliability}, 41(3):363--377, 1992.

\bibitem{fantechi2022wosar}
A.~Fantechi, G.~Gori, and M.~Papini.
\newblock Software rejuvenation and runtime reliability monitoring.
\newblock Submitted for publication.

\bibitem{gokhale2007}
S.~S. Gokhale.
\newblock Architecture-based software reliability analysis: Overview and
  limitations.
\newblock {\em IEEE Transactions on Dependable and Secure Computing},
  4(1):32--40, 2007.

\bibitem{he2021}
S.~He, P.~He, Z.~Chen, T.~Yang, Y.~Su, and M.~R. Lyu.
\newblock A survey on automated log analysis for reliability engineering.
\newblock {\em ACM Comput. Surv.}, 54(6), 2021.

\bibitem{hixenbaugh1968}
A.~F. Hixenbaugh.
\newblock Fault tree for safety.
\newblock Technical report, Boeing Aerospace Company, 1968.

\bibitem{huang1995software}
Y.~Huang, C.~Kintala, N.~Kolettis, and N.~D. Fulton.
\newblock Software rejuvenation: Analysis, module and applications.
\newblock In {\em Proc. of FTCS 1995}, pages 381--390, 1995.

\bibitem{iannino1990}
A.~Iannino and J.~D. Musa.
\newblock Software reliability.
\newblock In M.~C. Yovits, editor, {\em Advances in Computers}, volume~30,
  pages 85--170. Elsevier, 1990.

\bibitem{ieee2010_24765vocabulary}
{ISO/IEC/IEEE}.
\newblock {International Standard - Systems and software engineering --
  Vocabulary}.
\newblock {\em ISO/IEC/IEEE 24765:2010(E)}, pages 1--418, 2010.

\bibitem{lyu1996}
M.~R. Lyu.
\newblock {\em {Handbook of Software Reliability Engineering}}.
\newblock McGraw-Hill, Inc., 1996.

\bibitem{handbookRAMS2018}
Q.~Mahboob and E.~Zio.
\newblock {\em {Handbook of RAMS in Railway Systems: Theory and Practice}}.
\newblock {CRC Press}, 2018.

\bibitem{malhotra1994}
M.~{Malhotra} and K.~S. {Trivedi}.
\newblock Power-hierarchy of dependability-model types.
\newblock {\em IEEE Transactions on Reliability}, 43(3):493--502, 1994.

\bibitem{meyer1985}
J.~Meyer, A.~Movaghar, and W.~Sanders.
\newblock Stochastic activity networks: Structure, behavior, and application.
\newblock In {\em Proc. of Int. Workshop on Timed Petri Nets 1985}, pages
  106--115, 01 1985.

\bibitem{molloy1982}
M.~Molloy.
\newblock {Performance Analysis Using Stochastic Petri Nets}.
\newblock {\em IEEE Transactions on Computers}, 31(09):913--917, sep 1982.

\bibitem{moskowitz1958}
F.~Moskowitz.
\newblock The analysis of redundancy networks.
\newblock {\em Transactions of the American Institute of Electrical Engineers,
  Part I: Communication and Electronics}, 77(5):627--632, 1958.

\bibitem{nguyen2019_relAvalEval}
T.~A. Nguyen, D.~Min, E.~Choi, and T.~Dc Tran.
\newblock {Reliability and Availability Evaluation for Cloud Data Center
  Networks Using Hierarchical Models}.
\newblock {\em IEEE Access}, 7:9273--9313, 2019.

\bibitem{paolieri2019}
M.~Paolieri, M.~Biagi, L.~Carnevali, and E.~Vicario.
\newblock {The ORIS Tool: Quantitative Evaluation of Non-Markovian Systems}.
\newblock {\em IEEE Transactions on Software Engineering}, 47(6):1211--1225,
  2021.

\bibitem{papini_phd2021}
M.~Papini.
\newblock {\em {Reliability Evaluation of an Industrial System Through
  Predictive Diagnostics}}.
\newblock PhD thesis, Universit\`{a} degli Studi di Firenze, Florence, Italy, 6
  2021.

\bibitem{parlerCDE}
S.~G. Parler.
\newblock {Reliability of CDE Aluminum Electrolytic Capacitors, Cornell
  Dubilier white paper}, 2004.

\bibitem{pham2006}
H.~Pham.
\newblock {\em {System Software Reliability (Springer Series in Reliability
  Engineering)}}.
\newblock Springer-Verlag, Berlin, Heidelberg, 2006.

\bibitem{do254}
{RTCA and EUROCAE}.
\newblock {DO-254/ED-80 -- Design Assurance Guidance for Airborne Electronic
  Hardware}, 2000.

\bibitem{do178c}
{RTCA and EUROCAE}.
\newblock {DO-178C/ED-12C -- Software Considerations in Airborne Systems and
  Equipment Certification}, 2012.

\bibitem{arp4761}
{SAE International}.
\newblock {ARP4761 -- Guidelines and Methods for Conducting the Safety
  Assessment Process on Civil Airborne Systems and Equipment}, 1996.

\bibitem{arp4754a}
{SAE International}.
\newblock {ARP4754A -- Guidelines for Development of Civil Aircraft and
  Systems}, 2010.

\bibitem{stewart1994}
William~J. Stewart.
\newblock {\em Introduction to the Numerical Solution of Markov Chains}.
\newblock Princeton University Press, 1994.

\bibitem{relAvaEng2017}
K.~S. Trivedi and A.~Bobbio.
\newblock {\em {Reliability and Availability Engineering}}.
\newblock {Cambridge University Press}, 2017.

\bibitem{mil-hdbk-217}
{United States Department of Defense}.
\newblock {MIL-HDBK-217F: Reliability Prediction of Electronic Equipment},
  1991.

\bibitem{vaidyanathan1999}
K.~Vaidyanathan and K.S. Trivedi.
\newblock A measurement-based model for estimation of resource exhaustion in
  operational software systems.
\newblock In {\em Proc. of ISSRE 1999}, pages 84--93, 1999.

\bibitem{wang2007}
W.~Wang, W.~Xie, and K.~S. Trivedi.
\newblock Performability analysis of clustered systems with rejuvenation under
  varying workload.
\newblock {\em Performance Evaluation}, 64(3):247--265, 2007.

\bibitem{wood1996}
A.~{Wood}.
\newblock Predicting software reliability.
\newblock {\em Computer}, 29(11):69--77, 1996.

\end{thebibliography}

\end{document}